\documentclass{article}
\usepackage{amsfonts,amssymb,amsmath}
\usepackage{graphicx}
\newcommand{\R}{\mathbb{R}}

\begin{document}
\title{Hilbert Space Becomes Ultrametric in the High Dimensional Limit:
Application to Very High Frequency Data Analysis}

\author{Fionn Murtagh \\
Department of Computer Science, \\
Royal Holloway, University of London, \\
Egham, Surrey TW20 0EX, England. \\ 
fmurtagh@acm.org}

\maketitle

\begin{abstract}
An ultrametric topology formalizes the notion of hierarchical structure.
An ultrametric embedding, referred to here as ultrametricity, is 
implied by a natural hierarchical embedding.  Such hierarchical 
structure can be global in the data set, or local.  
By quantifying extent or degree of ultrametricity in a data set, we show
that ultrametricity becomes pervasive as dimensionality and/or
spatial sparsity increases.  This leads us to assert that very high
dimensional data are of simple structure.  We exemplify this finding through
a range of simulated data cases.  We discuss also application to very 
high frequency time series segmentation and modeling.
\end{abstract}

%\begin{keyword}      
%multivariate data analysis, cluster analysis, hierarchy, time series
%analysis, signal processing, ultrametric, p-adic
%\end{keyword}

\bigskip


%89.65.Gh Economics; econophysics, financial markets, business and management
%89.20.-a Interdisciplinary applications of physics
%02.50.-r Probability theory, stochastic processes, and statistics (see also 
%section 05 Statistical physics, thermodynamics, and nonlinear dynamical 
%systems)
%05.45.Tp Time series analysis

\section{Introduction}

The topology or inherent shape and form of an object is important.  In 
data analysis, the inherent form and structure of data clouds are important.
Quite a few models of data form and structure are used in data analysis.  
One of them is a hierarchically embedded set of clusters, -- a 
hierarchy.  It is traditional (since at least the 1960s) to impose such a
form on data, and if useful to assess the goodness of fit.  Rather than
fitting a hierarchical structure to data (e.g., \cite{rohlf}), 
our recent work
has taken a different orientation: we seek to find (partial or global) 
inherent hierarchical structure in data.  As we will describe in this 
article, there are interesting findings that result from this, and some
very interesting perspectives are opened up for data analysis and, potentially,
perspectives also on the physics (or causal or generative mechanisms) 
underlying the data.  

A formal definition of hierarchical structure is provided by ultrametric
topology (in turn, related closely  to p-adic number theory).  We will
return to this in section \ref{sect23} below.  First, though, we will 
summarize some of our findings.  

Ultrametricity is a pervasive property of observational data.  It 
arises as a limit case when data dimensionality or sparsity grows.  More
strictly such a limit case is a regular lattice structure and
ultrametricity is one possible representation for it.  Notwithstanding 
alternative representations, ultrametricity offers
 computational efficiency (related to tree depth/height being logarithmic 
in number of terminal nodes), linkage with dynamical or related 
functional properties (phylogenetic interpretation), and 
processing tools based on well known p-adic or ultrametric theory (examples: 
deriving a partition, or applying an ultrametric wavelet transform).  In 
\cite{khrenn} and other works, Khrennikov has pointed to the importance
of ultrametric topological analysis.  

Local ultrametricity is also of importance.  
This can be used for forensic data 
exploration (fingerprinting data sets):  see \cite{ref088} and \cite{ref0888}; 
and to expedite search and discovery in 
information spaces: see \cite{cha2} as discussed by us in \cite{ref08}, 
\cite{murpad}, and \cite{downs}. 

In section \ref{sect23} we show how extent of ultrametricity is measured.  
Section \ref{sect3} presents our main results on the remarkable properties
of very high dimensional, or very sparse, spaces.  As dimensionality 
or sparsity grow, so does the inherent hierarchical nature of the data 
in the space.  In section \ref{hfda} we then discuss application to 
very high frequency time series modeling.  

\section{Quantifying Degree of Ultrametricity}
\label{sect23}

Summarizing a full description in Murtagh \cite{ref08} we explored two
measures quantifying how ultrametric a data set is, -- Lerman's and a new
approach based on triangle invariance (respectively, 
the second and third approaches described in this section).

The triangular inequality holds for a metric space: $d(x,z) \leq 
d(x,y) + d(y,z)$ for any triplet 
of points $x,y,z$.  In addition the properties 
of symmetry and positive definiteness are respected.  The ``strong 
triangular inequality'' or ultrametric inequality is: $d(x,z) \leq 
\mbox{ max } \{ d(x,y), d(y,z) \}$ for any triplet $x,y,z$.  An
ultrametric
space implies respect for a range of stringent properties.  For
example, 
the triangle formed by any triplet is necessarily isosceles, with the
two
large sides equal; or is equilateral.

\begin{itemize}
\item 
Firstly, Rammal et al.\ \cite{ref09b} used discrepancy between each pairwise 
distance and the corresponding subdominant ultrametric.  Now, the
subdominant ultrametric is also known as the ultrametric distance
resulting from the single linkage agglomerative hierarchical
clustering method.   Closely related graph structures include the
minimal spanning tree, and graph (connected) components.  
While the subdominant provides a good fit to the
given distance (or indeed dissimilarity), it suffers from the
``friends of friends'' or chaining effect.  

\item
Secondly, Lerman \cite{ref07} developed a measure of ultrametricity,
termed H-classifiability,  using
ranks of all pairwise given distances (or dissimilarities).  The
isosceles (with small base) or equilateral requirements of the
ultrametric inequality impose constraints on the ranks.  The interval between
median and maximum rank of every set of triplets must be empty for 
ultrametricity.  We have used extensively  
Lerman's measure of degree of ultrametricity in a data set.  
Taking ranks provides scale invariance.  
But the limitation of Lerman's approach, 
we find, is that it is not reasonable to
study ranks of real-valued (values in non-negative reals) 
distances defined on a large set of points.

\item
Thirdly, our own measure of extent of ultrametricity \cite{ref08}
can be described algorithmically.  We 
examine triplets of points (exhaustively if possible, or otherwise
through sampling), and determine the three angles formed by the
associated triangle.  We select the smallest angle formed by the triplet
points.  Then we check if the other two remaining angles are
approximately equal.  If they are equal then our triangle is isosceles
with small base, or equilateral (when all triangles are equal).  The
approximation to equality is given by 2 degrees (0.0349 radians).  
Our motivation for
the approximate (``fuzzy'') equality is that it makes our approach
robust and independent of measurement precision.  
\end{itemize}

A supposition for use of our measure of ultrametricity is that we can can 
define angles (and hence triangle properties).  This in turn presupposes a
scalar product.  Thus we presuppose a complete 
normed vector space with a scalar product --
a Hilbert space -- to provide our needed environment.  

Quite a general way to 
embed data, to be analyzed, in a Euclidean space, is to use correspondence 
analysis \cite{ref08888}.  This explains our interest in using correspondence 
analysis: it provides a convenient and versatile 
way to take input data in many varied formats (e.g., ranks or scores, 
presence/absence,
frequency of occurrence, and many other forms of data) and map them into a 
Euclidean, factor space.

\section{Ultrametricity and Dimensionality}
\label{sect3}

\subsection{Distance Properties in Very Sparse Spaces}
\label{sect22}

Murtagh \cite{ref08}, and earlier work by Rammal et al.\ \cite{ref09a,ref09b},
has demonstrated the pervasiveness of ultrametricity, by 
focusing on the fact that  
sparse high-dimensional data tend to be ultrametric.  
In such work it is shown how numbers of points
in our clouds of data points are irrelevant; but what counts is the
ambient spatial dimensionality.  Among cases looked at are statistically 
uniformly
(hence ``unclustered'', or without structure in a certain sense)
distributed points, and statistically 
uniformly distributed hypercube vertices (so the
latter are random 0/1 valued vectors).  Using our ultrametricity
measure, there is a clear tendency to ultrametricity as the spatial
dimensionality (hence spatial sparseness) increases.  

%Figure \ref{fig6} illustrates these findings.  Dimensionality is increased
%up to 5000.   The number of points is kept at 5000, but varying this causes 
%little if any difference in the results.  (Exponentially increasing this 
%number of points, $n$, is quite a different matter, but is of no interest
%to us here.)  The results are quite similar irrespective of the 
%random number generation seed used, and also with respect to different 
%random generations of the data.  
%Two different sets of experiments are run, and the results 
%are displayed in the two curves.  Firstly, uniformly distributed real 
%values are
%used (hence $n$ points, in $\R^m$, where $m$ is the dimensionality), and 
%secondly uniformly distributed hypercube vertices are generated (hence
%$n$ points, in $ \{ 0, 1 \}^m$).  Using our triangle-based quantification of 
%ultrametricity, we find high dimensionality to approach the limit of 
%complete ultrametricity.  

%\begin{figure}
%\centering
%\includegraphics[width=9cm,angle=270]{fig-6-ultrametric.ps}
%\caption{Upper curve: uniformly distributed values.  Lower curve:
%random hypercube vertex points. A value of our triangle-based 
%ultrametricity measure equal to 1 is related
%to global ultrametricity.  For each dimensionality (50, 100, 500,
%1000, 5000) we used number of points $n = 5000$.  In other experiments
%we found very little variation as a function of $n$.}
%\label{fig6}
%\end{figure}

As \cite{hall} also show, Gaussian data behave in the same way and a 
demonstration of this is seen in Table \ref{tabunifgauss}.  
To provide an idea of consensus of these
results, the 200,000-dimensional Gaussian was repeated and yielded on 
successive runs values of the ultrametricity measure of: 0.96, 0.98, 0.96.  

\begin{table}
\begin{center}
\begin{tabular}{lllll} \hline
No. points &    Dimen. &  Isosc. &   Equil. &    UM   \\ \hline    
           &           &         &          &         \\
Uniform    &           &         &          &         \\
           &           &         &          &         \\
100        &    20     &    0.10 &    0.03  &    0.13 \\
100        &    200    &    0.16 &    0.20  &    0.36 \\
100        &    2000   &    0.01 &    0.83  &    0.84 \\
100        &    20000  &    0    &    0.94  &    0.94 \\
100        &    200000 &    0    &    0.97  &    0.97 \\
           &           &         &          &         \\
Hypercube  &           &         &          &         \\
           &           &         &          &         \\
100        &    20     &    0.14 &   0.02   &    0.16 \\
100        &    200    &    0.16 &   0.21   &    0.36 \\
100        &    2000   &    0.01 &   0.86   &    0.87 \\
100        &    20000  &    0    &   0.96   &    0.96 \\
100        &    200000 &    0    &   0.97   &    0.97 \\  
           &           &         &          &         \\
Gaussian   &           &         &          &         \\
           &           &         &          &         \\
100        &    20     &    0.12 &    0.01  &    0.13 \\
100        &    200    &    0.23 &    0.14  &    0.36 \\
100        &    2000   &    0.04 &    0.77  &    0.80 \\
100        &    20000  &    0    &    0.98  &    0.98 \\
100        &    200000 &    0    &    0.96  &    0.96 \\ \hline
\end{tabular}
\end{center}
\caption{Typical results, based on 300 sampled triangles from triplets of
points.  For uniform, the data are generated on [0, 1]; hypercube vertices
are in $\{ 0, 1\}^{\mbox{Dimen}}$, 
and for Gaussian, the data are
of mean 0, and variance 1.  Dimen. is the ambient
dimensionality.  Isosc. is the number of isosceles triangles with 
small base, as a
proportion of all triangles sampled.  Equil. is the number of equilateral 
triangles as a proportion of triangles sampled.  UM is the proportion of 
ultrametricity-respecting triangles (= 1 for all ultrametric).}
\label{tabunifgauss}
\end{table}

In the following, we explain why high dimensional and/or sparsely populated
 spaces are ultrametric.  

As dimensionality grows, so too do distances (or indeed
dissimilarities, if
they do not satisfy the triangular inequality).  The least change
possible for dissimilarities to become distances has been formulated
in terms of the smallest additive constant needed, to be added to all
dissimilarities \cite{tor,cai1,cai2,neu}.
Adding a sufficiently large 
constant to all dissimilarities transforms them into a set of
distances.  Through addition of a larger constant, it follows that
distances become approximately equal, thus verifying a trivial case of
the ultrametric or ``strong triangular'' inequality.  Adding to
dissimilarities or distances may be a direct consequence of increased
dimensionality. 

For a close fit or good approximation,  
the situation is not as simple for taking dissimilarities, or
distances,
into ultrametric distances.  A best fit solution is given by \cite{desoete}
(and software is available in R \cite{hornik}).
If we want a close fit to the given 
dissimilarities then a good choice would avail either of the maximal 
inferior, or subdominant, ultrametric; or the minimal superior
ultrametric.
Stepwise algorithms for these are commonly known as, respectively,
single
linkage hierarchical clustering; and complete link hierarchical
clustering.  
(See \cite{ref2,ref07,mur85a} and other texts on 
hierarchical clustering.)  

\subsection{No ``Curse of Dimensionality'' in Very High Dimensions}
\label{sect21}

Bellman's \cite{bel}   ``curse of dimensionality'' relates to exponential 
growth of hypervolume as a function of dimensionality.  Problems
become tougher as dimensionality increases.  In particular problems 
related to proximity search in high-dimensional spaces tend to become
intractable.  

In a way, a ``trivial limit'' (Treves \cite{ref120}) 
case is reached as dimensionality increases.  This makes high
dimensional 
proximity search very different, and given an appropriate data
structure -- such as a binary hierarchical clustering tree -- we can
find nearest neighbors in worst case $O(1)$ or constant computational
time \cite{ref08}.  The proof is simple: the tree data structure
affords a constant number of edge traversals.  

The fact that limit properties are ``trivial'' makes them no less 
interesting to study.  Let us refer to such ``trivial'' properties as
(structural or geometrical) regularity properties (e.g.\ all points
lie on a regular lattice).  

First of all, the symmetries of regular
structures in our data may be of importance.  For example, processing 
of such data can exploit these regularities.  

Secondly, ``islands'' or
clusters in our data, where each ``island'' is of regular structure,
may be of interpretational value.  

Fourthly, and finally, regularity of particular properties does not 
imply regularity of all properties.  So, for example, we may have only 
partial existence of pairwise linkages.  

\begin{figure}
\includegraphics[width=16cm]{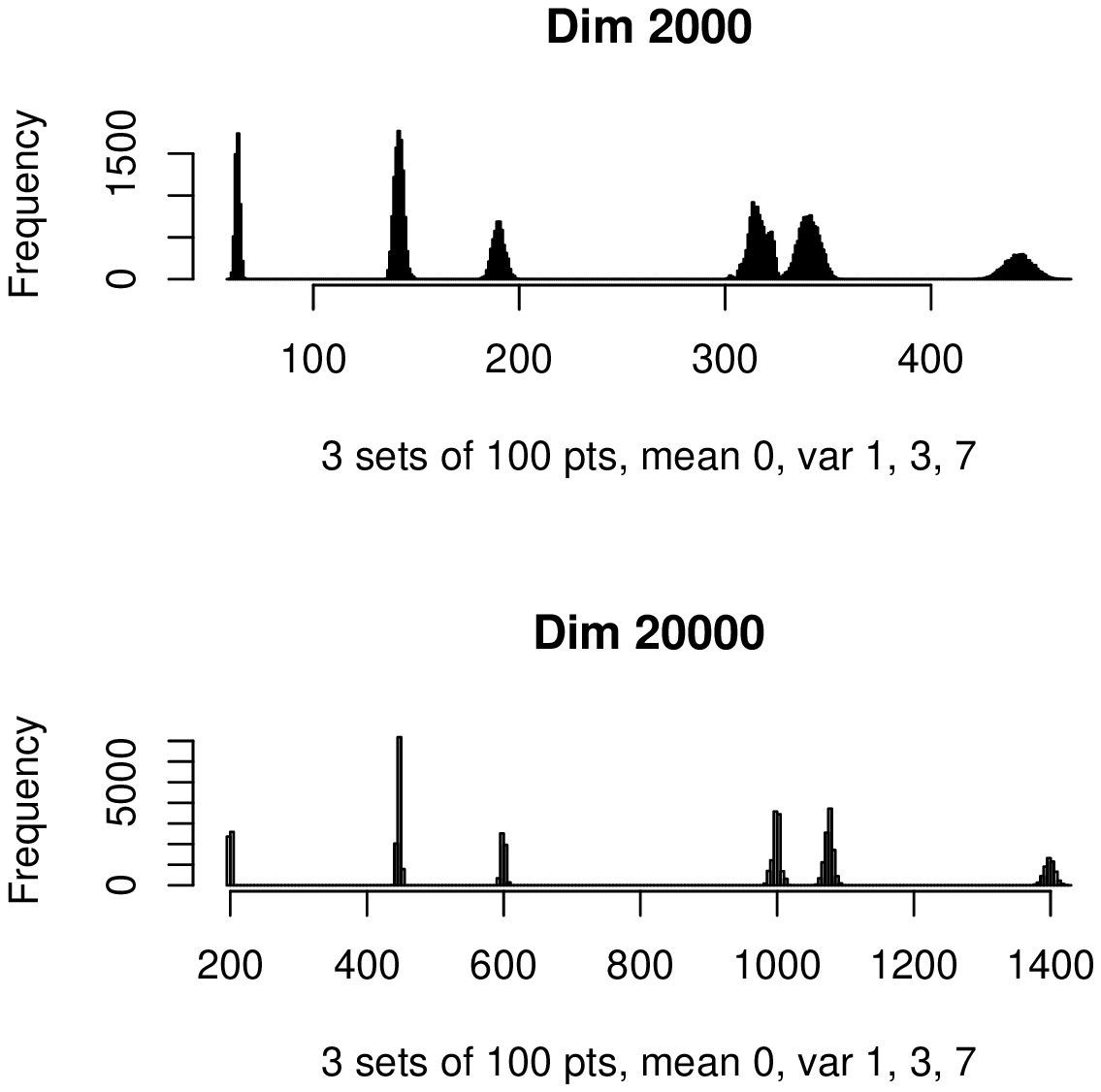}
\caption{An illustration of how ``symmetry'' or ``structure'' can become 
increasingly pronounced as dimensionality increases.
Shown are two simulations, each with 3 sub-populations of 
Gaussian-distributed data,
in, respectively, ambient dimensions of 2000 and 20,000.  These simulations
correspond to the 3rd last, and 2nd last, rows of Table \ref{tabunifgauss}.}
\label{fig1}
\end{figure}

Thus we see that in very high dimensions, and/or in very (spatially) 
sparse data clouds, 
there is a simplification of structure, 
which can be used to mitigate any ``curse of dimensionality''.  
Figure \ref{fig1}
shows how the distances within and between clusters become tighter with 
increase in 
dimensionality.  

\subsection{Gaussian Clusters in Very High Dimensions}
\label{sect33}

\subsubsection{Introduction}

We will distinguish between cluster characteristics as follows:

\begin{enumerate}
\item cluster size: number of points per cluster;
\item cluster location: here, mean, identical on every dimension;
\item cluster scale: here, standard deviation, identical on 
every dimension.
\end{enumerate}

These cluster characteristics are simple ones, and future work will 
consider greater sophistication.

In the 
homogeneous clouds studied in Table \ref{tabunifgauss} it is seen that 
the isosceles (with small base) case disappeared early on, as dimensionality
increased greatly, to the advantage of the equilateral case of ultrametricity.
So the points become increasingly equilateral-related as dimensionality 
grows.  This is not the case when the data in clustered, as we will now
see.

\subsubsection{Clusters with Different Locations, Same Scale}

\begin{figure}
\includegraphics[width=16cm]{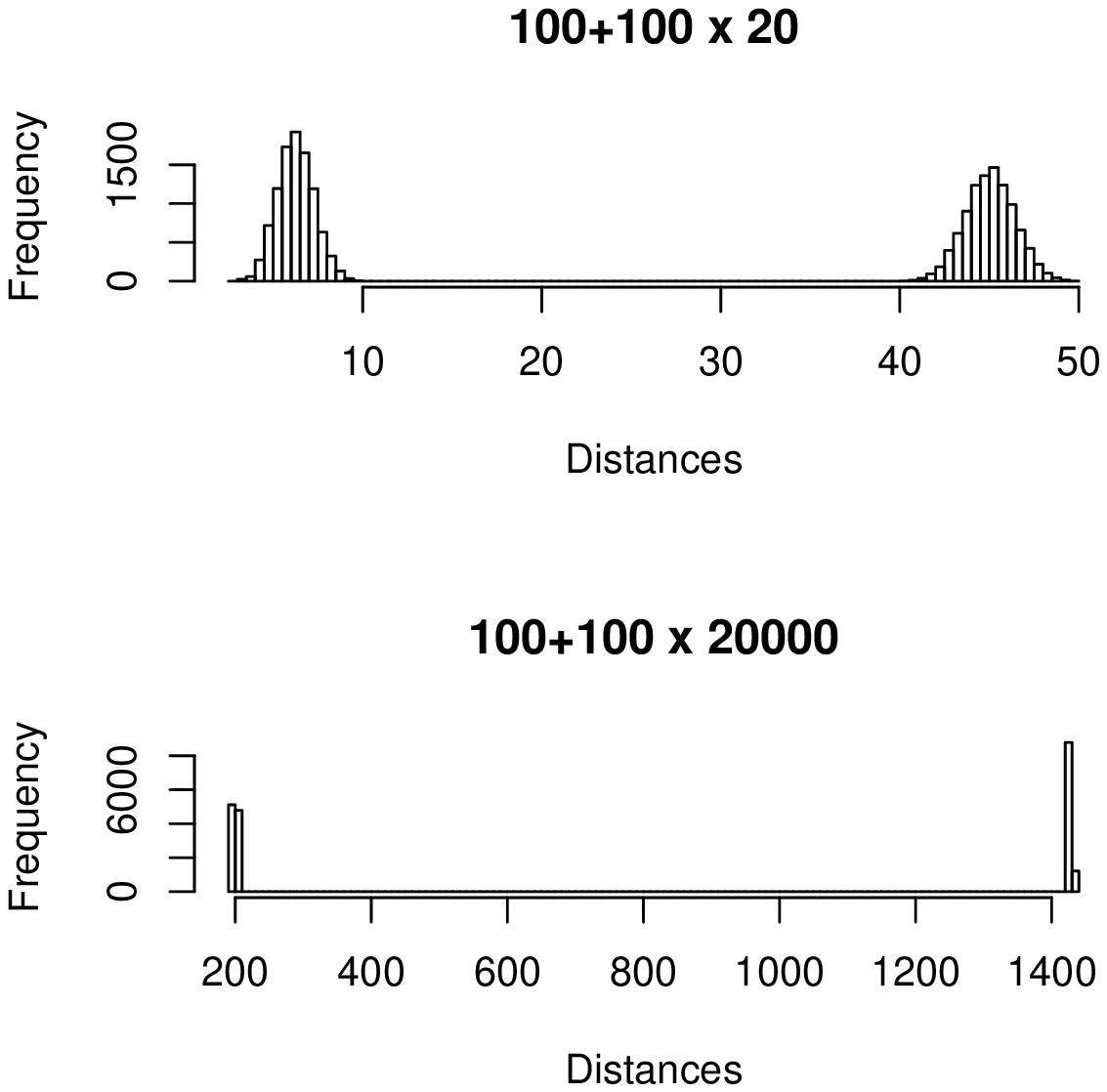}
\caption{A further 
illustration of how ``symmetry'' or ``structure'' can become 
increasingly pronounced as dimensionality increases, relating to the 
$200 \times 20$ and $200 \times 20,000$ (first of the succession of rows)
cases
of Table \ref{tabgauss}.  These are histograms of all interpoint
distances, based on two Gaussian clusters.  The first has mean 0 and standard
deviation 1 on all dimensions.  The second has mean 10 and standard deviation 
 1 on all dimensions.}  
\label{fig2}
\end{figure}

Table \ref{tabgauss} is based on two clusters, and shows how isosceles
triangles increasingly dominate as dimensionality grows.  Figure \ref{fig2}
illustrates low and high dimensionality scenarios relating to Table
\ref{tabgauss}.  There is clear confirmation in this table as to how 
interrelationships in the cluster become more compact and, in a certain 
sense, more trivial, in high dimensions.  This does not obscure the 
fact that we indeed have hierarchial relationships becoming ever more 
pronounced as dimensionality, and hence relative sparsity, increase. 
These observations help us to see quite clearly just how hierachical 
relationships come about, as ambient dimensionality grows.  

\begin{table}
\begin{center}
\begin{tabular}{lllll} \hline
No. points &    Dimen. &  Isosc. &   Equil. &    UM   \\ \hline    
           &           &         &          &         \\
200        &    20     &    0.08 &    0  &    0.08 \\
200        &    200    &    0.19 &    0.04  &    0.23 \\
200        &    2000   &    0.42 &    0.20  &    0.62 \\
200        &    20000  &    0.74  &    0.22  &    0.96 \\ 
           &           &          &          &         \\
200        &    20000  &    0.7   &    0.28  &    0.98 \\ 
200        &    20000  &    0.77  &    0.21  &    0.98 \\ 
200        &    20000  &    0.76  &    0.21  &    0.98 \\ 
200        &    20000  &    0.75  &    0.24  &    0.99 \\ 
200        &    20000  &    0.73  &    0.25  &    0.98 \\ \hline
\end{tabular}
\end{center}
\caption{Results based on 300 sampled triangles from triplets of
points.  Two Gaussian clusters, each of 100 points,  were used in each case.
One point set was of mean 0, and the other of mean 10, on each dimension.
The standard deviations on each dimension were 1 in all cases.  Column 
headings are as in Table \ref{tabunifgauss}.  Five further results are given
for the 20,000-dimension case to show variability.}
\label{tabgauss}
\end{table}

\subsubsection{Clusters with Different Locations, Different Scales}

\begin{figure}
\includegraphics[width=16cm]{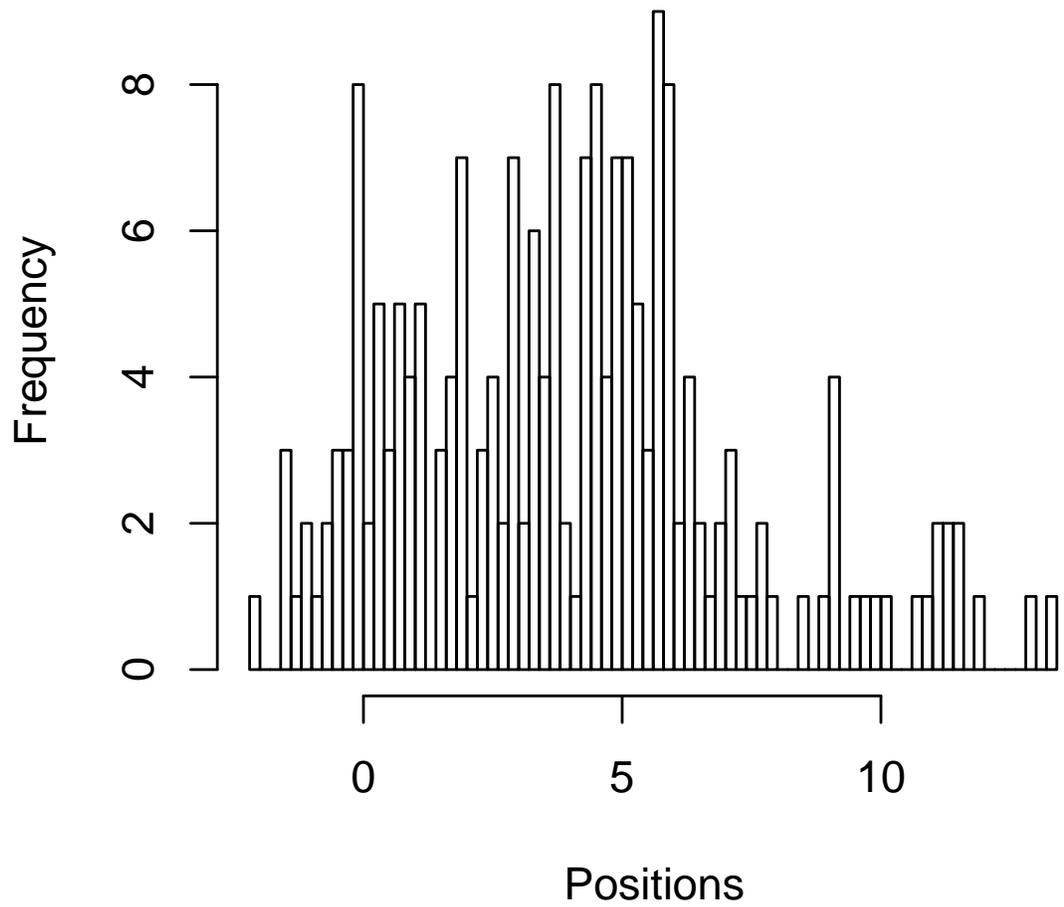}
\caption{A projection onto one dimension, to illustrate the less than
clearcut clustering problem addressed.  There are four Gaussians here,
each of 50 realizations, with means at 0, 3, 5 and 8, and with respective
standard deviations of 1, 2, 1, 3.}
\label{fig4}
\end{figure}

A more demanding case study is now tried.  We generate 50 points per cluster
with the following characteristics: mean 0, standard deviation 1, on each 
dimension; mean 3, standard deviation 2, on each dimension; mean 5, standard
deviation 1, on each dimension; and mean 8, standard deviation 3, on each 
dimension.  Table \ref{tabgauss2} shows the results obtained.  Here 
we have not achieved quite the same level of ultrametricty, due to slower
growth in ultrametricity which is, in turn, due to the more murky, 
less dermarcated, but undoubtdely clustered, set of data.  Figure 
\ref{fig4} illustrates this: this histogram shows one dimension, where 
we note that means of the Gaussians are at 0, 3, 5 and 8.  

\begin{table}
\begin{center}
\begin{tabular}{lllll} \hline
No. points &    Dimen. &  Isosc. &   Equil. &    UM   \\ \hline    
           &           &         &          &         \\
200        &    20     &    0.04 &    0.01  &    0.05 \\
200        &    200    &    0.11 &    0.05  &    0.16 \\
200        &    2000   &    0.28 &    0.06  &    0.34 \\
200        &    20000  &    0.5  &    0.08  &    0.58 \\
200        &    200000 &    0.55 &    0.11  &    0.66 \\ \hline
\end{tabular}
\end{center}
\caption{Results based on 300 sampled triangles from triplets of
points.  Four Gaussian clusters, each of 50 points,  were used in each case.
See text for details of properties of these clusters.} 
\label{tabgauss2}
\end{table}

\begin{figure}
\includegraphics[width=16cm]{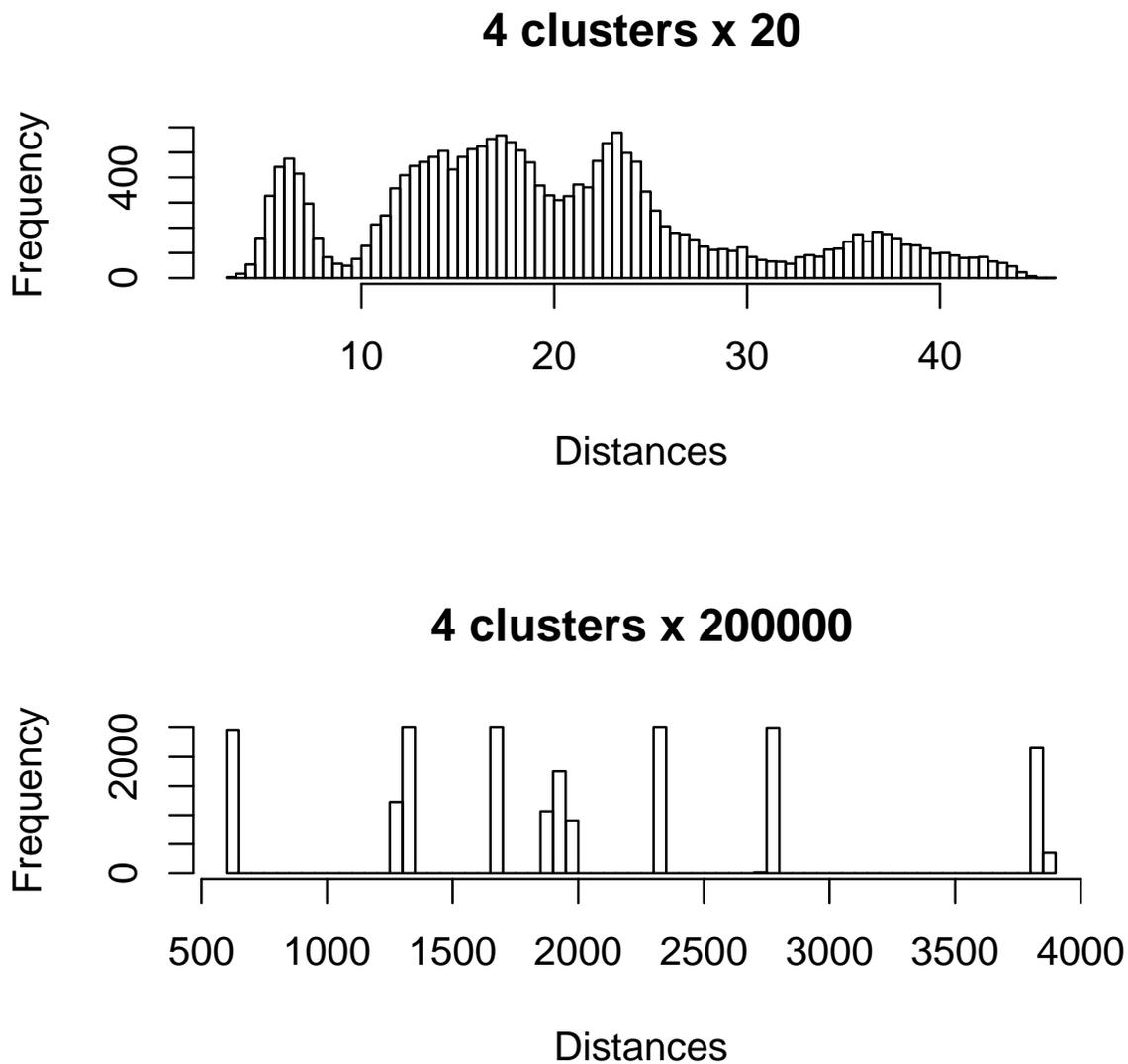}
\caption{Compaction of distances with rise in dimensionality: 4 clusters,
substantially overlapping are the basis for the histograms of all 
pairwise distances. Top: ambient dimensionality 20.  Bottom: ambient 
dimensionality 200,000.}
\label{fig3}
\end{figure}

When we look closer at Table \ref{tabgauss2}, as shown in Figure 
\ref{fig3}, the compaction of distances is again very interesting.  
We verified the 7 peaks found in the lower histogram in Figure \ref{fig3},
and available but confusedly overlapping and ill-defined in the upper 
histogram of Figure \ref{fig3}.

What we find for the 7 peaks is as follows.  
Distances within the clusters correspond to: peaks 1,
2, 3 and (again) 1.  That two clusters are associated with one peak is
clear from the fact that two of our clusters are of identical scale.  

We can examine inter-cluster distances and we found these to be associated
with peaks: 2, 3, 4, 5, 6, 7.  Given 4 clusters, we could well have up to 
6 possible additional peaks.  

%Distances within cluster 2: peak 2.
%Distances within cluster 3: peak 3.
%Distances within cluster 4: peak 1.
%Distances between clusters 1 and 2, and within clusters 1 and 2: 
%peaks 1, 2 and 3.
%Distances between clusters 1 and 3, and within clusters 1 and 3: 
%peaks 1, 3 and 7.
%Distances between clusetrs 1 and 4, and within clusters 1 and 4: 
%peaks 1 and 5.
%Distances between clusters 2 and 3, and within clusters 2 and 3: 
%peaks 2, 3 and 6.
%Distances between clusters 2 and 4, and within clusters 2 and 4: 
%peaks 1 and (subdivided) 2.
%Distances between clusters 3 and 4, and within clusters 3 and 4: 
%peaks 1, 3 and 4. 
%
%By a process of eliminating the known single cluster intra-cluster
%distances, we find: 
%inter 1,2    peak 3
%inter 1,3    peak 7
%inter 1,4    peak 5
%inter 2,3    peak 6
%inter 2,4    differnented peak 2
%inter 3,4    peak 4

%In general terms, we find that peaks 1, 2 and 3 are based on intra-cluster
%distances, which is quite reasonable given that there are three different
%standard deviations used to generate the clusters.  We find then 
%that peaks 4, 5, 6 and 7 are based on iter-cluster distances.  

\subsubsection{Conclusions on High Dimensional Gaussian Clouds}

From these case studies, it is clear that increased dimensionality 
sharpens and distinguishes the clusters.  
If we can embed data -- any data -- in a far
higher ambient dimensionality, without destroying the interpretable 
relationships in the data, then we can so much more easily read off the 
clusters.  

To read off clusters, including memberships and properties, our findings
can be summarized as follows.  

For cluster size (i.e., numbers of points per cluster), sampling alone 
can be used, and we do not pursue this here.  

For cluster scale (i.e., standard deviation, assumed the same on each 
dimension), we associate each cluster, or a pair of clusters, with each
peak.    The total number of peaks gives an upper bound on the number of
clusters.  (For $k$ clusters, we have $ \leq k + k \cdot (k-1) /2$ 
peaks.) 

Using cluster scale also permits use of the following cluster model: 
suppose that all clusters are defined to have intra-cluster distance 
that is less than inter-cluster distance.  Then it follows that the 
peaks of lower distance correspond to the clusters (as opposed to pairs
of clusters).  

An example of this is as follows.  
In Figure \ref{fig3}, lower panel, we read from left to 
right, applying the following algorithm: select the first $k$ peaks as 
clusters, and ask: are there sufficient peaks to represent all 
inter-cluster pairs?  If we choose $k = 3$, there remain 4 peaks, which is
too many to account for the inter-cluster pairs (i.e., $ 3 \cdot (3-1)/2)$).  
So we see that Figure \ref{fig3} is incompatible with 
$k = 3$ or the presence of just 3 clusters.  

Consequently we move to $k = 4$, and see that Figure \ref{fig3} is 
consistent with this.  

A further identifiability assumption is reasonable albeit not required:
that all smallest peaks be associated with intra-cluster distances.  
This need not be so, since we could well have a dense cluster superimposed on
a less dense one.  However it is a reasonable parimony assumption.  Supported
by this assumption, Figure \ref{fig3} points to a minimum of 4 clusters in 
the data, with up to 4 peaks (read off from left to right, i.e., in increasing
order of distance) corresponding to these clusters.

\section{Applications}

\subsection{Data Recoding in the Correspondence Analysis Tradition}

The iris data has been very widely used as a toy data set since 
Fisher used it in 1936 (\cite{fish}, taken from \cite{anderson}) 
to exemplify discriminant analysis.  It consists of 150 iris flowers, 
each characterized by 4 petal and sepal, width and breadth, measurements.  
On the one hand, therefore, we have the 150 irises in $\R^4$.  Next, 
each variable value was recoded by us 
to be a rank (all ranks of a given variable
considered) and the rank was boolean-coded (viz., for the top rank variable 
value, $1000 \dots$, for the second rank variable value, $0100 \dots$, etc.).
Following removal of zero total columns, the second data set 
defined the 150 irises in $\R^{123}$.  Actually, this definition of the 
150 irises is in fact in $\{0,1\}^{123}$.  

Our triangle-based measure of the degree of ultrametricity in a data set
(here the set of irises), with 0 = no ultrametricity, and 1 = every triangle
an ultrametric-respecting one, gave the following: for irises in 
$\R^4$, 0.017; and for irises in $\{0,1\}^{123}$: 0.948.  
% See G4, UMexpts 

This provides a nice illustration of how recoding can dramatically change 
the picture provided by one's data.  In chapter 3 of \cite{ref08888} it is
discussed just what change in the data cloud is caused by the recoding.  
Our objective here is not to pursue the goodness of fit or otherwise of 
one data encoding vis-\`a-vis another.  Instead our objective is to point
out how data encoding influences directly (and at times remarkably) the
data cloud's ultrametricity, or ease of being hierarchically embedded.  

In correspondence analysis, the $\chi^2$ distance when 
used on data tables with constant
marginal sums becomes a weighted Euclidean distance.  This is 
important for us as data analyst, 
because it means that we can directly influence the 
analysis by equi-weighting, say, the table rows in the following way: 
we double the row vector values by including an absence (0 value) 
whenever there is a presence (1 value) and vice versa.  Or for a table of 
percentages, we take both the original value $x$ and $100 - x$.  In the 
correspondence analysis tradition \cite{ref2,ref08888} this is known as 
{\em doubling} ({\em d\'edoublement}).  
More generally, booleanizing, or making qualitative, data in this way, for a
varying (value-dependent) number of target value categories (or modalities) 
leads to the form of coding known as {\em complete disjunctive form}.  

Such coding increases the embedding dimension, and data sparseness.
From our example of recoding the Fisher data, such coding 
can influence degree of ultrametricity.  We conclude that careful 
data coding can increase the extent to which our data is inherently 
hierarchical.  Furthermore the latter in turn may be beneficial in 
enhancing data interpretability (for example, by unravelling phylogenetic
aspects expressed by the data).

\subsection{Application to High Frequence Data Analysis}
\label{hfda}

In this section we establish proof of concept for application of 
the foregoing work to analysis of very high frequency 
univariate time series signals.  

Consider each of the cases considered in section \ref{sect33},
expressed there as $n \times m$ arrays, as instead representing $n$ segments,
each of (contiguous) length $m$, of a time series or one-dimensional signal.
Assuming our aim is to cluster these segments on the basis of their 
properties, then it is reasonable to require that they be non-overlapping.  
The $n$ segments could come from anywhere, in any order, in the time series.
So for the case of an $n \times m$ array considered previously, then 
implies a time series of length at least $n m$.  The most immediate way 
to construct the time series is to raster scan the $n \times m$ array, 
although alternatives come readily to mind.  

The methodology discussed in section \ref{sect33} then is seen to be also
a time series segmentation approach, facilitating the characterizing of 
the segments used.  

\begin{figure}
\includegraphics[width=16cm]{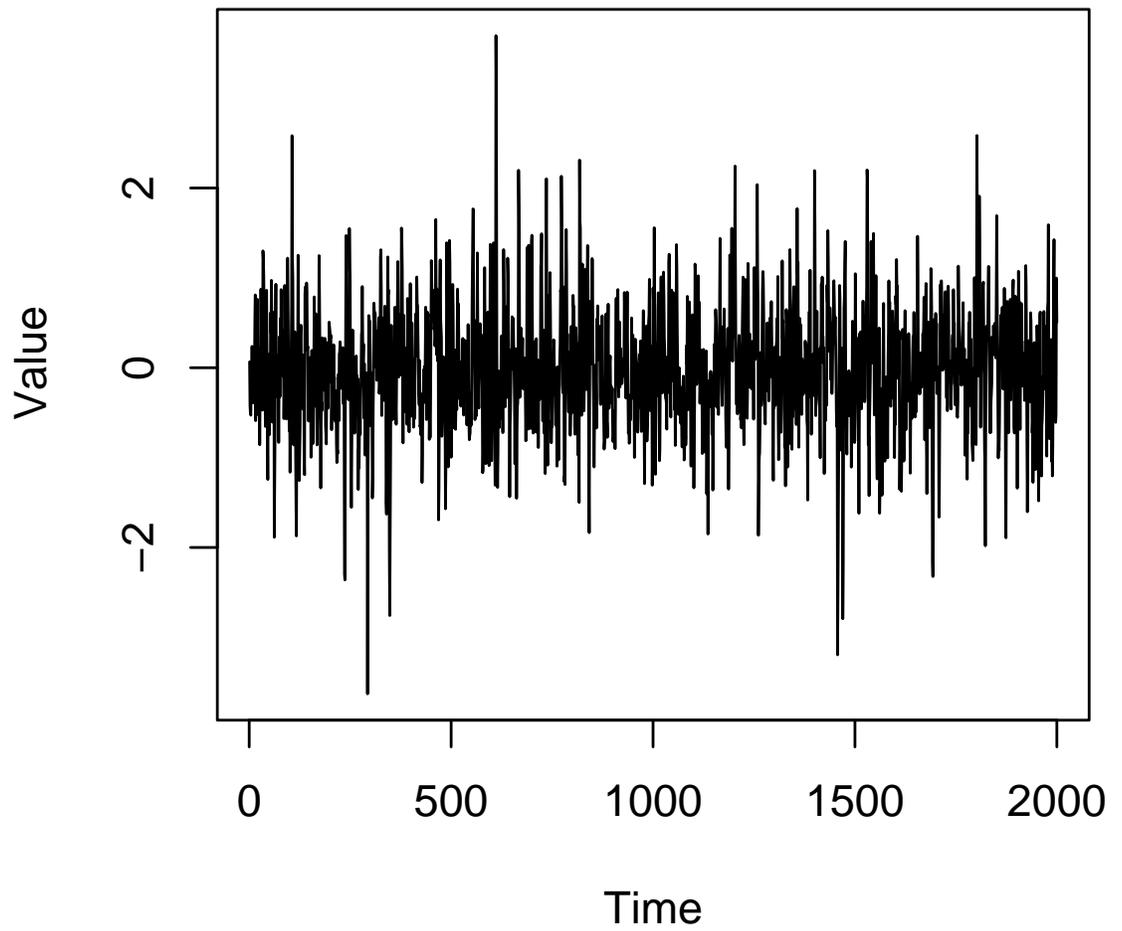}
\caption{Sample (using first 1000 values) of a time series segment, based
on the first ARIMA set of parameters. (Order 2 
AR parameters: $0.8897, -0.4858$,
MA parameters: $-0.2279, 0.2488$.)}
\label{fig6}
\end{figure}

\begin{figure}
\includegraphics[width=16cm]{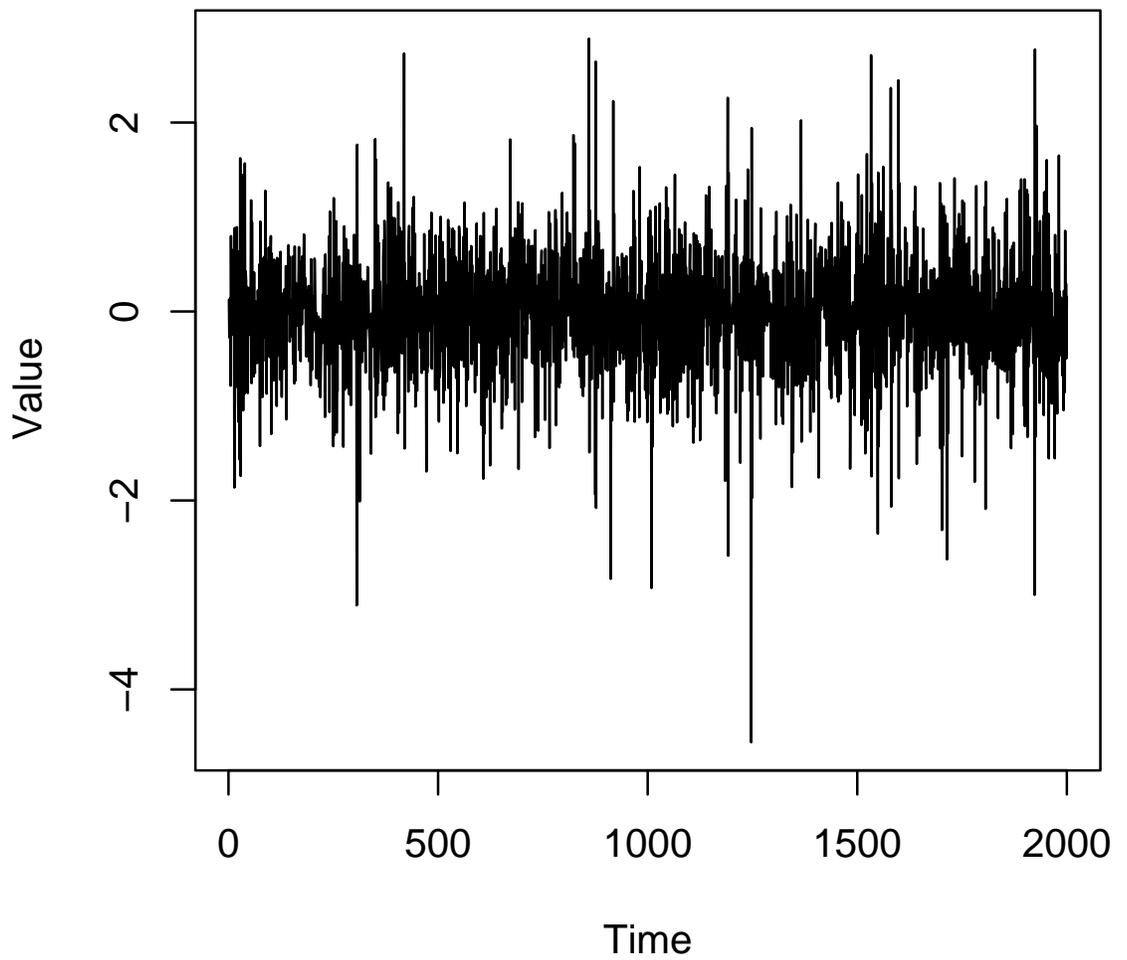}
\caption{Sample (using first 1000 values) of a time series segment, based
on the second ARIMA set of parameters. (Order 2
AR parameters: $0.2897, -0.1858$,
MA parameters: $-0.7279, 0.7488$.)}
\label{fig7}
\end{figure}

\begin{figure}
\includegraphics[width=16cm]{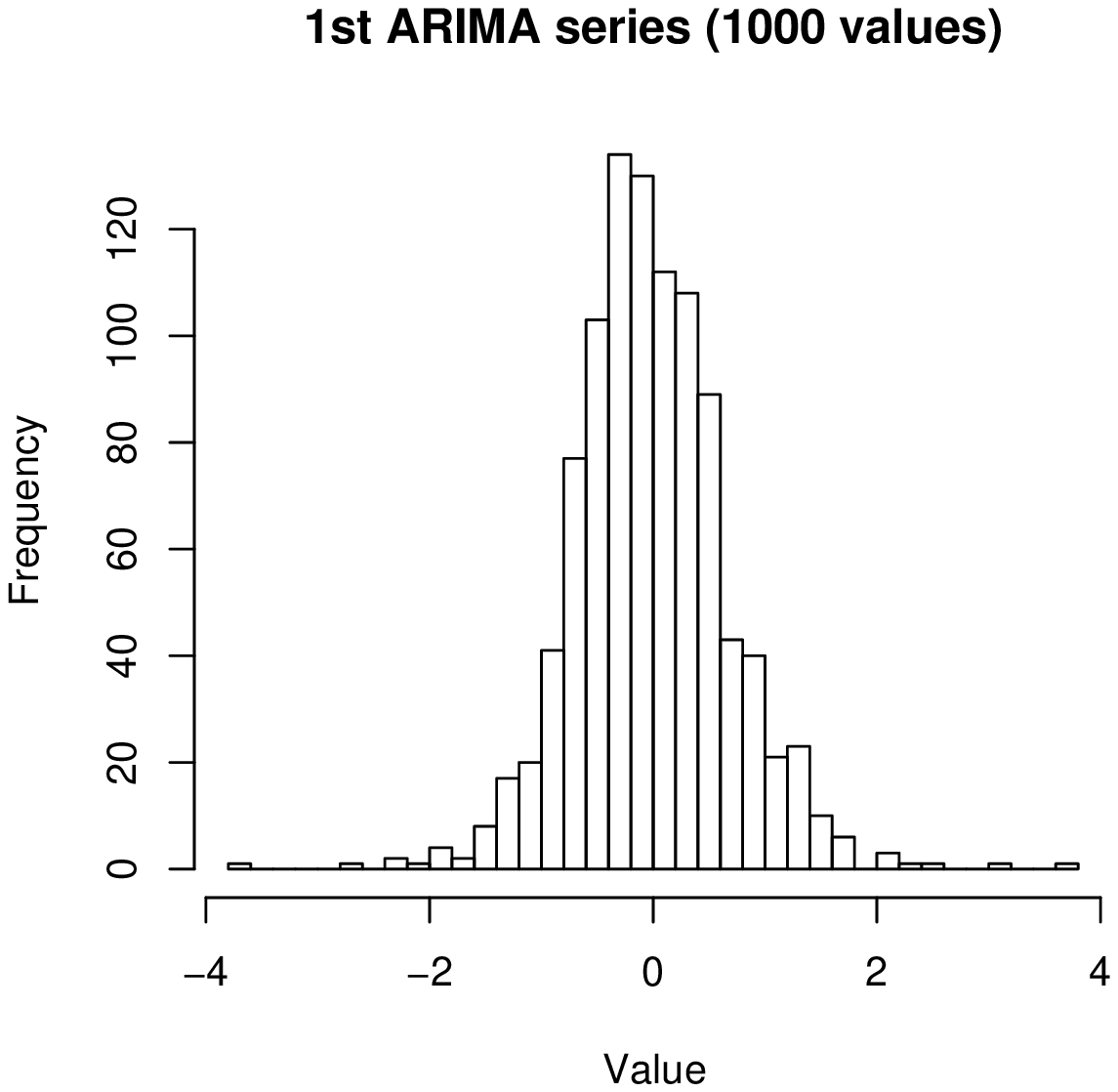}
\caption{Histogram of sample (using first 1000 values) of 
time series segment shown in Figure \ref{fig6}.}
\label{fig8}
\end{figure}

\begin{figure}
\includegraphics[width=16cm]{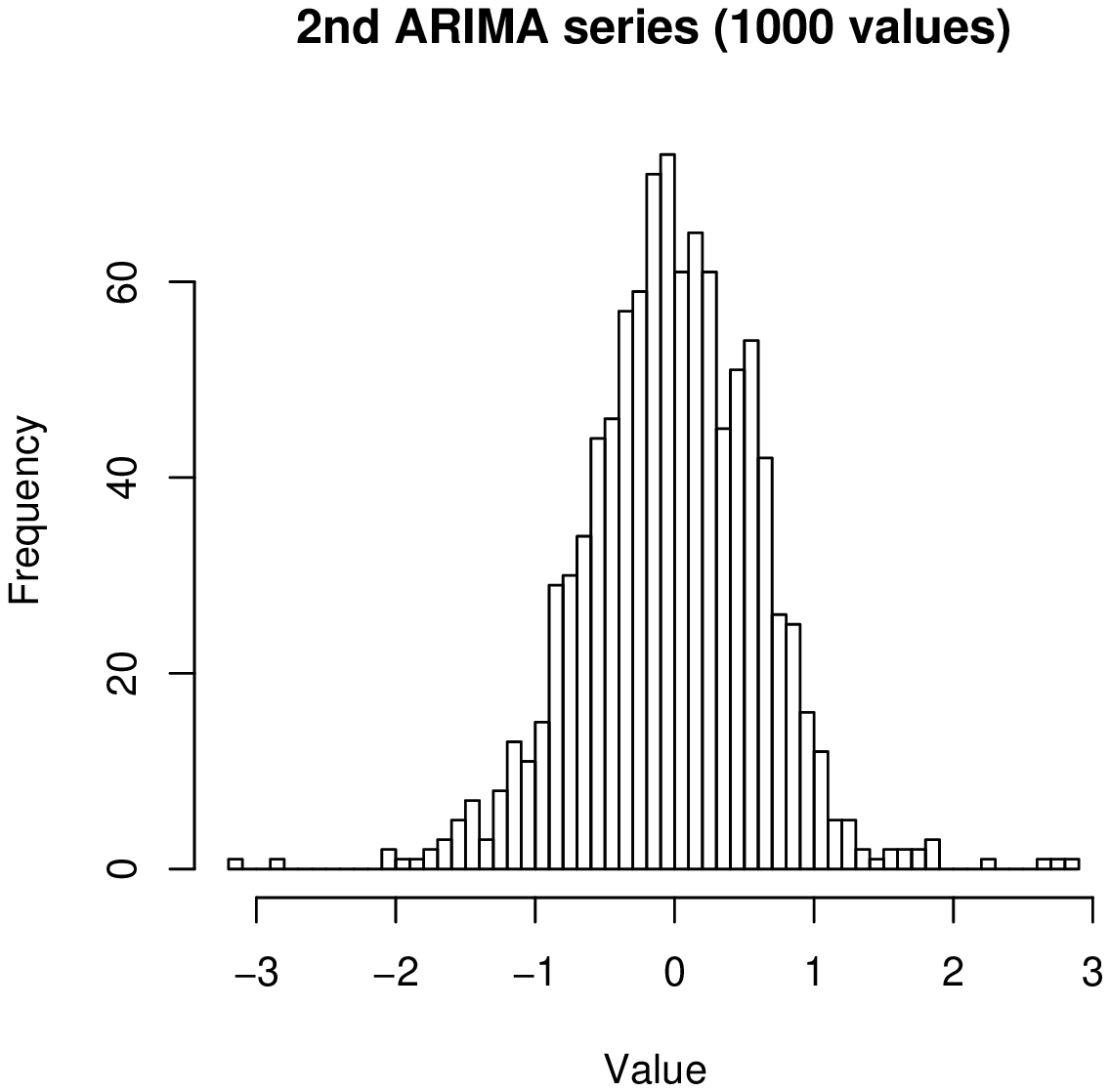}
\caption{Histogram of sample (using first 1000 values) of 
time series segment shown in Figure \ref{fig7}.}
\label{fig9}
\end{figure}

To explore this further we consider a time series consisting of two 
ARIMA (autoregressive integrated moving average) models, with parameters:
order, autoregression coefficients, moving average coefficients, and a 
``mildly longtailed'' set of innovations based on the Student t distribution 
with 5 degrees of freedom.  % The R function arima.sim was used.  
Figures \ref{fig6} and \ref{fig7} show samples of these time series 
segments.  Figures \ref{fig8} and \ref{fig9} show histograms of these 
samples. 

Table \ref{tabfindat} shows typical results obtained in regard to 
ultrametricity.  The dimensionality can be considered as the embedding
dimension.  Here, although ultrametricity increases, and the 
equilateral configuration seems to be increasing but with decrease of the 
isosceles with small base configuration, we do not consider it of 
practical relevance to test with even higher ambient dimensionalities.  
It is clear from the data, especially Figures \ref{fig8} and \ref{fig9},
that the two signal models are very close in their properties.  

\begin{figure}
\includegraphics[width=16cm]{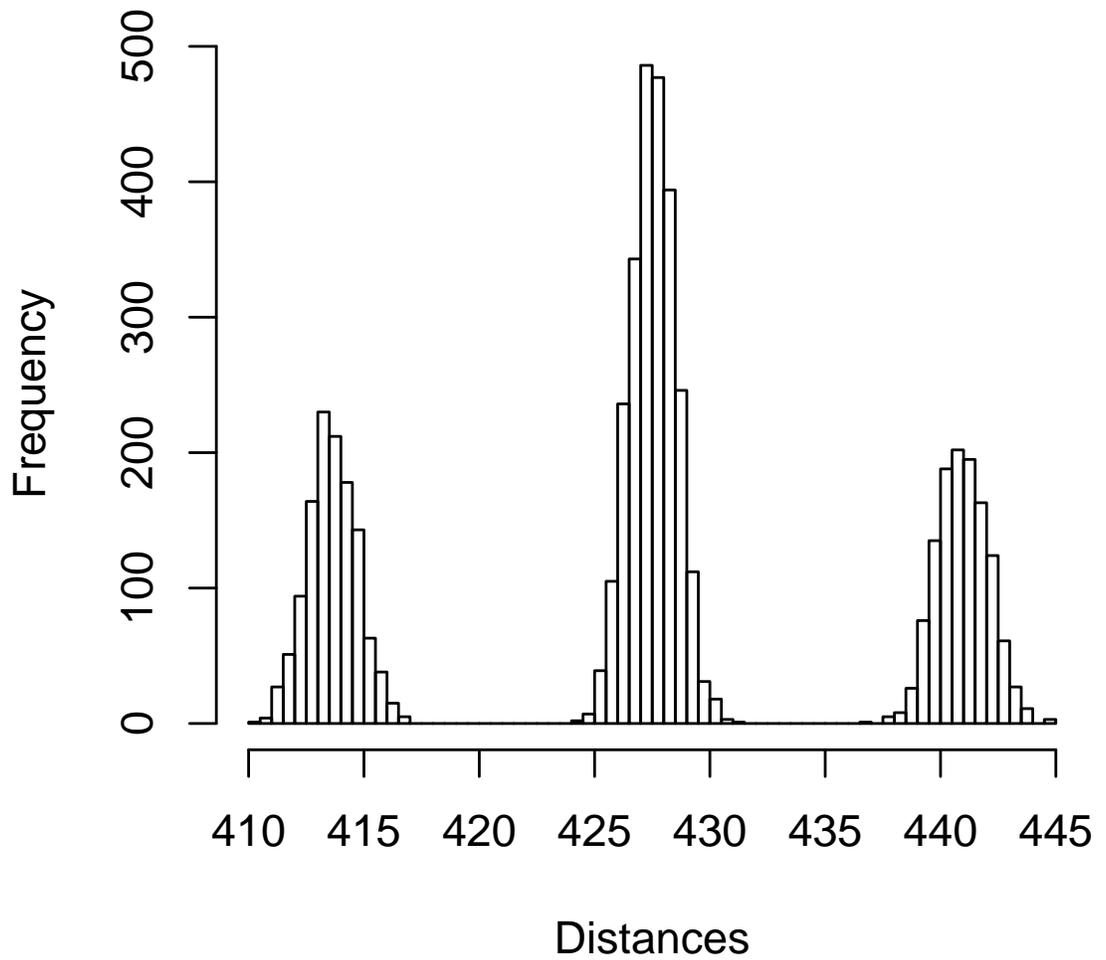}
\caption{Histogram of distances from 100 time series segments, 
using 50 segments each from the two ARIMA models, and using an 
embedding dimensionality of 200,000.}
\label{fig10}
\end{figure}

Examining the histograms of all inter-pair time series segments, both 
intra and inter cluster, we find the clearly distinguished peaks shown 
in Figure \ref{fig10}.  As before, we use Euclidean distance between 
time series segments or vectors.  (We note that normalization or other 
transformation is not particularly relevant here.  In fact we want to 
distinguish between inter and intra cluster cases.  Furthermore the 
unweighted Euclidean distance is consistent with our use of angles to 
quantify triangle invariants, and hence respect for ultrametricity properties.)

\begin{table}
\begin{center}
\begin{tabular}{lllll} \hline
No. time series &    Dimen. &  Isosc. &   Equil. &    UM   \\ \hline    
           &           &         &          &         \\
100        &    2000   &    0.17 &    0.32  &    0.49 \\
100        &    20000  &    0.15  &    0.5  &    0.65 \\
100        &    200000 &    0.03 &    0.57  &    0.60 \\ \hline
\end{tabular}
\end{center}
\caption{Results based on 300 sampled triangles from triplets of
points.  Two sets of the ARIMA models are used, each of 50 realizations.}
\label{tabfindat}
\end{table}

We find clearly distinguishable peaks in Figure \ref{fig10}.  The lower and 
the higher peaks belong to the two ARIMA components.  The central peak 
belongs to the inter-cluster distances.  

We have shown that our methodology can be of use for time series segmentation
and for model identifiability.  We will assess this further in future 
work.  Given the use of a Hilbert space as the essential springboard of 
all aspects of this work, it would appear that generalization of this work 
to multivariate time series analysis is straightforward.  What remains 
important, however, is the availability of very large embedding 
dimensionalities, i.e.\ very high frequency data streams.

\section{Conclusions}

What we have observed in all of 
this work is that in the limit of high dimensionality 
a Hilbert space becomes ultrametric.  

It has been our aim in this work to link observed data with an ultrametric
topology for such data.  The traditional approach in data analysis, of course, 
is to impose structure on the data.  This is done, for example, by using 
some agglomerative hierarchical clustering algorithm.  We can always 
do this (modulo distance or other ties in the data).  Then we can assess
the degree of fit of such a (tree or other) structure to our data.  
For our purposes, here, this is unsatisfactory.  

\begin{itemize}
\item Firstly, our aim was to show 
that ultrametricity can be naturally present in our data, globally or 
locally.  We did not want any ``measuring tool'' such as an 
agglomerative hierarchical clustering algorithm to overly influence 
this finding.  (Unfortunately \cite{ref09b} suffers from precisely this 
unhelpful influence of the ``measuring tool'' of the subdominant 
ultrametric.  In other respects, \cite{ref09b} is a seminal paper.) 

\item
Secondly, let us assume that we did use hierarchical clustering, and then 
based our discussion around the goodness of fit.  This again is a traditional
approach used in data analysis, and in statistical data modeling.  But such 
a discussion would have been unnecessary and futile.  For, after all, if 
we have ultrametric properties in our data then many of the widely used
hierarchical clustering algorithms will give precisely the same outcome, 
and furthermore the fit is by definition optimal.  

\end{itemize}
%In linking data with an ultrametric embedding, whether local only, or global,
%we have, in this article, 
%proceeded also in the direction of exploiting this achievement.  
%While some applications, like discrimination between time series signals, 
%or texts, have been covered here, other applications like bioinformatics 
%database search and discovery, and analysis of large scale 
%cosmological structures \cite{murpad}, 
%have just been opened up.  In \cite{ezhov} this methodology is applied
%to quantum statistics. 

We have described an application of this work to very high frequency 
signal processing.  The twin objectives are signal segmentation, and model 
identification.  We have noted that a considerable amount of this work is 
model-based: we require assumptions (on clusters, and on model(s)) for 
identifiability.  

Motivation for this work includes the availability of very high frequency 
data streams in various fields (physics, 
engineering, finance, meteorology, 
bio-engineering,  and bio-medicine).  By using a very large embedding 
dimensionality, we are approaching the data analysis on a very gross 
scale, and hence furnishing a particular type of multiresolution analysis.  
That this is worthwhile has been shown in our case studies.

\end{document}